# Non-negligible Contributions to Thermal Conductivity From Localized Modes in Amorphous Silica

Wei Lv[1], Asegun Henry[*,1,2]

**Affiliations:**

[1]George W. Woodruff School of Mechanical Engineering, Georgia Institute of Technology, Atlanta, GA 30332, USA.

[2]School of Materials Science and Engineering, Georgia Institute of Technology, Atlanta, GA, 30332, USA.

[*]Corresponding author: ase@gatech.edu

**Abstract**

Thermal conductivity is an important property for almost all applications involving heat transfer, ranging from energy and microelectronics to food processing and textiles. The theory and modeling of crystalline materials is in some sense a solved problem, where one can now calculate the thermal conductivity of any crystalline line compound from first principles [1,2] using expressions based on the phonon gas model (PGM) [3,4]. However, modeling of amorphous materials still has many open questions, because the PGM itself becomes questionable when one cannot rigorously define the phonon velocities. New theories and methods are therefore needed to understand phonon transport in amorphous materials. In this letter, we used our recently developed Green-Kubo modal analysis (GKMA) method to study amorphous silica (a-$SiO_2$). The predicted thermal conductivities exhibit excellent agreement with experiments at all temperatures and anharmonic effects are included in the thermal conductivity calculation for all types of modes in a-$SiO_2$ for the first time. Previously, localized modes (locons) have been thought to have a negligible contribution to thermal conductivity, due to their highly localized nature, which conceptually should prevent them from moving energy to another location. However, in a-$SiO_2$ our results indicate that locons contribute more than 10% to the total thermal conductivity from 400K to 800K and they are largely responsible for the increase in thermal conductivity of a-$SiO_2$ above room temperature. This is an effect that cannot be explained by previous methods and therefore offers new insight into the nature of phonon transport in amorphous/glassy materials.

Understanding the underlying physics of thermal conductivity（TC）in crystalline dielectric materials has made a great progress in recent years [1,5–10]. Using expressions based on the phonon gas model (PGM) and first principles calculations, it is now possible to predict the TC of dielectric crystalline line compounds very accurately [11–13]. Furthermore, the PGM has guided our understanding of phonon transport in micro and nano-structures and has provided a more than satisfactory explanation of the behavior in a variety of systems [14–16]. As a result, it has rightfully become our primary lens for interpreting phonon transport related phenomena.

From the PGM perspective, the lattice TC is described by the phonon contributions to the specific heat, the phonon group velocities and mean free paths (MFPs) [17,18]. The specific heat can be calculated from a lattice dynamics (LD) calculation of the phonon frequencies [18,19]. The phonon velocities can also be obtained from LD calculations of the dispersion $\omega(k)$, where the group velocity is given by $d\omega/d\mathbf{k}$. The MFP is the product of the group velocity and time between phonon scattering events (e.g., the relaxation time), which can be calculated from normal mode analysis or first principle calculations [2,20]. For crystalline line compounds, or ordered alloys, the system is periodic, which allows for straightforward definition of the dispersion and thus the group velocity is well defined. However for amorphous materials, or other structurally/compositionally disordered systems, due to the lack of periodicity, one cannot clearly define the group velocity [21]. As a result, since the PGM hinges on knowledge of the velocities, application of the PGM to amorphous materials is highly questionable. Several studies [22,23] have related the behavior of low frequency modes in amorphous materials (e.g., propagons) to propagating modes in crystalline materials, because they exhibit plane-wave modulated vibrations. The majority of the modes in amorphous materials, however, are still non-propagating modes (e.g., diffusons and locons) [24] and as a result, MFP based explanations of phonon transport in amorphous materials become difficult to rationalize.

In light of this issue, the seminal work of Allen and Feldman (A-F) [25], which expressed the TC of amorphous materials based on a mode diffusivity instead of MFP has therefore become widely accepted in understanding the phonon transport in amorphous materials, since it does not require one to define the phonon velocity. However, its limitation is that it does not include the effects of anharmonicity in the atomic interactions. Thus, it is interesting to consider what might be the effects of anharmonicity on the contributions of different modes in material such as a-SiO$_2$, for which the A-F method does not exhibit satisfactory agreement [26]. A-SiO$_2$ is also of widespread technological significance and therefore understanding its thermal transport physics is of significant and broad interest. Of particular importance is the fact that the TC of a-SiO$_2$ increases with temperature beyond room temperature, which is somewhat difficult to explain with MFP based arguments. The model developed by A-F, for example, predicts essentially constant TC above 200 K [26]. Thus, it is of interest to consider the role anharmonicity might play in understanding the TC of a-SiO$_2$.

Recently Lv and Henry [21] developed a new method termed the Green-Kubo Modal Analysis (GKMA) method which is a combination of Green-Kubo (GK) formula with modal information from LD. The GKMA method is significant because it incorporates all degrees of anharmonicity, since it uses molecular dynamics (MD) simulations to obtain the time history of the modal contributions to the heat current operator. Applying the GKMA method to crystalline silicon (c-Si) and amorphous silicon (a-Si) agrees well with other modal analysis methods and experimental data, which has confirmed its viability and accuracy in describing the mode level details of phonon transport in both crystalline and glassy materials.

Here, we briefly summarize the GKMA formalism so that its key features can be emphasized. First, the normal mode coordinates/shapes are computed from a LD calculation of the entire atomic supercell of the material in question. Then, by projecting the atomic velocities from a MD simulation of the same supercell onto the normal mode coordinates, one can obtain the time history of each normal mode's amplitude. Each atom's instantaneous velocity can then be decomposed into individual mode contributions based on the respective instantaneous normal mode amplitudes, whereby summing the modal contributions returns the atom velocity. One then substitutes the modal components of each atom's velocity into the heat flux operator [21] to obtain each mode's instantaneous contribution to the heat flux. The total heat flux can then be obtained from the sum of all individual mode contributions to the heat flux, via

$$\mathbf{Q} = \sum_{n}^{3N} \mathbf{Q}(n) = \sum_{n}^{3N} \frac{1}{V} \sum_{i} \left[ E_i \mathbf{v}_i(n) + \sum_{j} (-\nabla_{\mathbf{r}_i} \Phi_j \cdot \mathbf{v}_i(n)) \mathbf{r}_{ij} \right] \quad (1)$$

where n is mode number, $N$ is total number of atoms in the super cell, $V$ is volume of the super cell, $E_i$ is the kinetic and potential energy of atom i, $\Phi_j$ denotes potential energy of atom j, $\mathbf{v}_i(n)$ is the mode velocity of mode n on atom i. and $\mathbf{r}_{ij}$ is distance between atom i and j. Now having access to the individual mode heat fluxes, we can substitute the summation over modes in Eq. (1) directly into the Green-Kubo expression for TC, which describes the TC as proportional to the heat flux autocorrelation function. One then obtains the TC as a direct summation over individual mode contributions,

$$\kappa = \sum_{n} \kappa(n) = \sum_{n} \frac{V}{k_B T^2} \int_0^\infty \langle \mathbf{Q}(n,t) \cdot \mathbf{Q}(0) \rangle dt \quad (2)$$

where $\kappa(n)$ stands for TC contribution of mode n, $k_B$ is Boltzmann constant, T is the temperature and V is volume. One can substitute the mode heat flux into the second instance of the heat flux autocorrelation function to obtain a double summation over mode-mode cross-correlations, which provides immense insight into the interrelationships and interactions between modes as will be shown later (in Fig. 4). Using Eq. (2) one can calculate the TC of individual modes in any material where the atoms vibrate around stable equilibrium sites, using one unified formalism. Classical MD has been considered to be inaccurate at low temperatures (below Debye Temperature) because it does not reproduce the proper mode amplitudes that correspond to the quantum occupations. As a result, classical MD results in a constant heat capacity with temperature, since every mode is equally excited at all temperatures. However, once each individual mode's TC is obtained, one can easily apply a quantum specific heat correction, which allows one to extend the MD based predictions to essentially any temperature. To obtain the most accurate temperature dependence, one can use the following expression,

$$\kappa(T) = \sum_n f_Q(\omega(n,T), T) \cdot f_\kappa(n,T) \tag{3}$$

which includes three explicit functions of temperature, namely $f_Q$, $f_\kappa$ and $\omega$. In Eq. (3) the function $f_Q$ represents the ratio of the quantum to classical specific heat for mode n, which has frequency $\omega$ at temperature T and is unit less. The function $f_\kappa$ represents the GKMA derived mode TC contributions (e.g., it has units of TC), obtained from MD simulations at discrete values of T. The function $\omega$ represents the phonon frequency of mode n, which itself also exhibits some temperature dependence. Later, in Fig. 2, we show how TC prediction improves as each aspect of the temperature dependence is included, which will henceforth be referred to using the subscripts $Q$, $\kappa$, and $\omega$.

The first and most important source of temperature dependence is in the quantum to classical specific heat ratio $f_Q$, which is what causes the TC to decrease to zero as T→0K. Furthermore, it restricts the contributions of the high frequency modes at low temperatures and modulates the MD derived TC contributions determined from the GKMA method. The second important source of temperature dependence enters through the GKMA derived TC contributions $f_\kappa$. As temperature changes, the modal interactions change and the contributions of different modes are inherently temperature dependent via the anharmonic nature of the interactions. However, unlike the quantum specific heat correction, which is a continuous function of temperature, MD simulations are run at discrete temperatures. To then generate a piece-wise continuous function for TC vs. temperature, one can interpolate the data for $f_\kappa$ at discrete values of temperature. Here, one can use the data at a few initial temperatures and determine by inspection what temperature ranges may require additional simulations to improve the resolution of the temperature dependence in temperature ranges where the contributions change more rapidly. It is advantageous to minimize the number of temperatures needed for $f_\kappa$ to minimize computational expense. Lastly, the phonon frequencies ($\omega$) change with temperature, via thermal expansion (e.g., frequency softening). The extent of the frequency shift as a function of temperature can be determined by interpolation of the data at discrete temperatures, using the peak frequency obtained from a Fourier transform of the mode amplitudes.

Using the GKMA method, we then calculated the modes and their respective contributions to TC for a-SiO$_2$. All details associated with the calculations are given in the supplementary materials. The inverse participation ratio (IPR) and phonon density of states (DOS) are shown in Fig. 1 (a) and (b). The IPR quantifies the extent of localization for a given mode [24]. From Fig.

1(b), there are two regions that have localized modes, from 25 to 30 THz and above 35 THz. Given their higher IPR, we classified both of these groups of modes as locons [24], which are spatially localized and typically only involve a small group of atoms in the vibration (examples of locon normal mode shapes are given in supplementary information). Interestingly, the modes between 30-35 THz are diffusons, which are delocalized over the entire system. Thus, a-$SiO_2$ appears to exhibit two regions of locons [26], which is distinctly different from a-Si [24]. Figure 1(c) shows the TC accumulation with frequency at different temperatures. The accumulations above 400K are very similar, which indicates that the anharmonic effects do not drastically change the mode-mode interactions above 400K.

Using Eq. (3), one can also use the GKMA derived TC contributions with a quantum specific heat correction [21] $f_Q$, to extend the applicability of the calculations to any temperature. Figure 1(c) and (d) shows the TC accumulation vs. frequency before and after quantum specific heat corrections at five different temperatures for a-$SiO_2$. Figure 2 shows how the predictions improve, as more accurate temperature dependent information is included. Initially the TC is calculated directly from classical MD, and as expected, the temperature dependence qualitatively differs from the experimental data [27] as shown in Fig. 2 (a). The data in Fig. 2(a) corresponds to evaluating Eq. (3) and setting $f_Q = 1$, for all modes. However, after the quantum specific heat correction is applied to GKMA results, the overall experimental trend is obtained. However, the results still differ significantly if only the GKMA accumulation is used for a single temperature. Figure 2(b) corresponds to evaluating Eq. (3) with $f_Q$ equal to the quantum to classical specific heat ratio, but the temperature dependence of both $f_\kappa$ and $\omega$ are neglected, as we have only used the values of $f_\kappa$ at 400K from GKMA and harmonic frequencies $\omega_0$ at 0K from LD. In reality, the accumulation, which is obtained from the TC contributions $f_\kappa$, itself is a function of temperature, as indicated by Fig 1(c). The accumulation, however, exhibits moderate and rather monotonic temperature dependence, which is well approximated by linear interpolation at a few key temperatures. Thus, once the quantum correction ($Q$) and accumulation ($\kappa$) temperature dependence are applied, the agreement with experiments improves significantly. We then correct for the temperature dependence of the phonon frequencies ($\omega$) themselves (e.g., softening), which can be determined from a Fourier Transform of each mode's kinetic energy [28]. This softening tends to shift the frequencies lower by ~ 10% at 800K and 6% at 400K, which is important, because the quantum correction is sensitive to the mode frequencies. In Fig 2(c), the TC predictions shift to the left once the frequency softening $\omega(n,T)$ is incorporated into the calculations. In Fig. 2, we have used the subscripts $Q$, $\kappa$, and $\omega$ to denote the quantum specific heat temperature dependence, TC contribution temperature dependence and mode frequency dependence respectively. GKMA fully includes anharmonicity resulting in different quantitative predictions than the A-F method, which computes the mode diffusivity based on harmonic approximation [25] and assumes the diffusivity is not temperature dependent. Instead of using temperature independent thermal diffusivity, we interpolated the values at several temperatures to incorporate the temperature dependence of $f_\kappa$.

After all the temperature dependence and interpolations have been accounted for, the magnitude and trend of the temperature dependent TC exhibits excellent agreement with experiments as shown in Fig 2 (d). We thus recommend that all three temperature dependence should be included to obtain the most accurate predictions using Eq. 3. In Fig. 2, we have included the uncertainty associated with sampling a limited number of ensembles, which is given as the standard deviation of the GK results at that temperature. Comparing with other methods, GKMA demonstrates much better agreement with experiments for a-SiO$_2$ as shown in Fig 3(a), which derives from the more complete inclusions of mode dependence, anharmonicity and its temperature dependence.

One of the most striking features of Fig. 1 is the contributions associated with locons. Previously, locons have been thought to exhibit negligible contributions to TC, due to their restricted spatial extent, which renders it difficult to imagine how they can transfer a significant amount of heat. Nonetheless, despite the long held belief that locons are negligible, Fig. 1 shows that they contribute significantly to the TC of a-SiO$_2$. Most notably, even after quantum correction, locons are responsible for approximately half of the continual rise in TC for a-SiO$_2$ above room temperature. Furthermore, the cross-correlation maps at different temperatures of Fig. 4 show that even though the autocorrelation is dominant in at all temperatures, as temperature increases, the magnitude of cross-correlation contribution increases. This behavior suggests that modes interact more strongly with other modes of differing frequency as temperature increases, which is consistent with our intuitive understanding of anharmonicity. The locon contributions come from both cross-correlations and auto-correlations, but it remains unclear how these modes are able to help transfer energy from one location to another. Previous GKMA study of a-Si [21] shows that locons have negligible contribution to the TC. Allen and Feldman have used harmonic approximation and reached a similar conclusion for a-Si that locons do not contribute to the TC [25]. After their work on a-Si, researchers started to generally neglect the locon contributions to the TC of amorphous materials. However, using GKMA method for a-SiO$_2$, we find that locon contributions are non-negligible. In Fig. 3(b), the dashed blue curve represents the TC contributions from locons and the solid blue curve is the TC vs. temperature with the locon contribution neglected. Without the contributions from locons, the TC begins to deviate from the experiments after 250K.

We have applied our recently developed GKMA method to a-SiO$_2$, and have incorporated three sources of temperature dependence. Our results indicate that in order to obtain the most accurate predictions, one should incorporate a quantum correction on the heat capacity, as well as the temperature dependence of the GKMA thermal conductivity contributions and also the softening of the mode frequencies themselves. With these effects included, we have demonstrated the best agreement with experiments as compared to all previous models. Further examination of the contributions from different types of modes revealed the first quantitative evidence that localized modes (e.g., locons) can contribute significantly to the total TC of a material. In a-SiO$_2$, locons are responsible for approximately half of the rise in TC above 400K. Further study is needed to develop a new physical picture that can describe how both locons and extended, but non-propagating modes such as diffusons are able to transfer heat through disordered materials.

## Acknowledgments

This research was supported Intel grant AGMT DTD 1-15-13 and computational resources were provided by the Partnership for an Advanced Computing Environment (PACE) at the Georgia Institute of Technology and National Science Foundation supported XSEDE resources (Stampede) under grant numbers DMR130105 and TG- PHY130049.

**Figures**

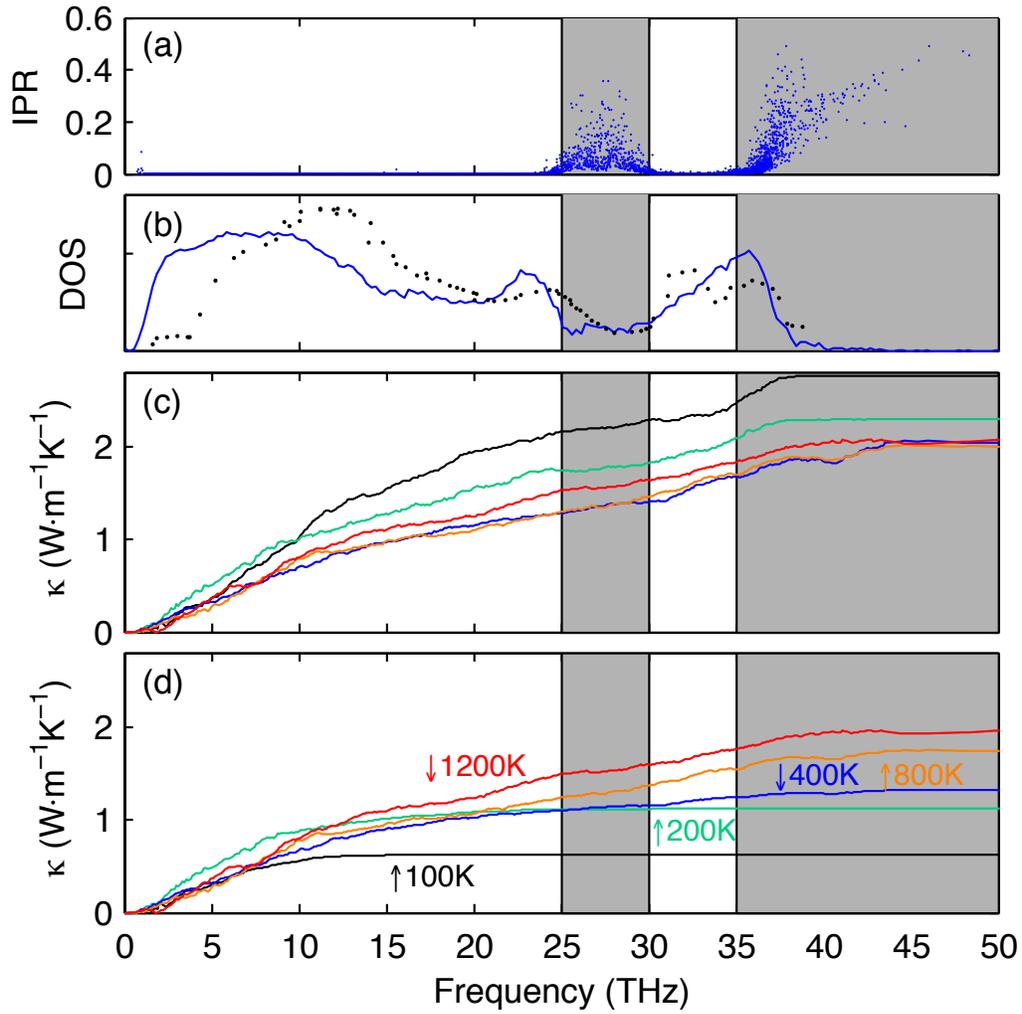

FIG. 1. (a) IPR of modes in a-SiO$_2$; (b) Phonon density of states (solid blue curve) and experimental results (black circles) [29]; (c) TC accumulation vs. mode frequency for a-SiO$_2$ using GKMA at different temperatures (100K, 200K, 400K, 800K, 1200K) w/o quantum specific heat correction; (d) with quantum specific heat correction. The gray shaded areas represent locons.

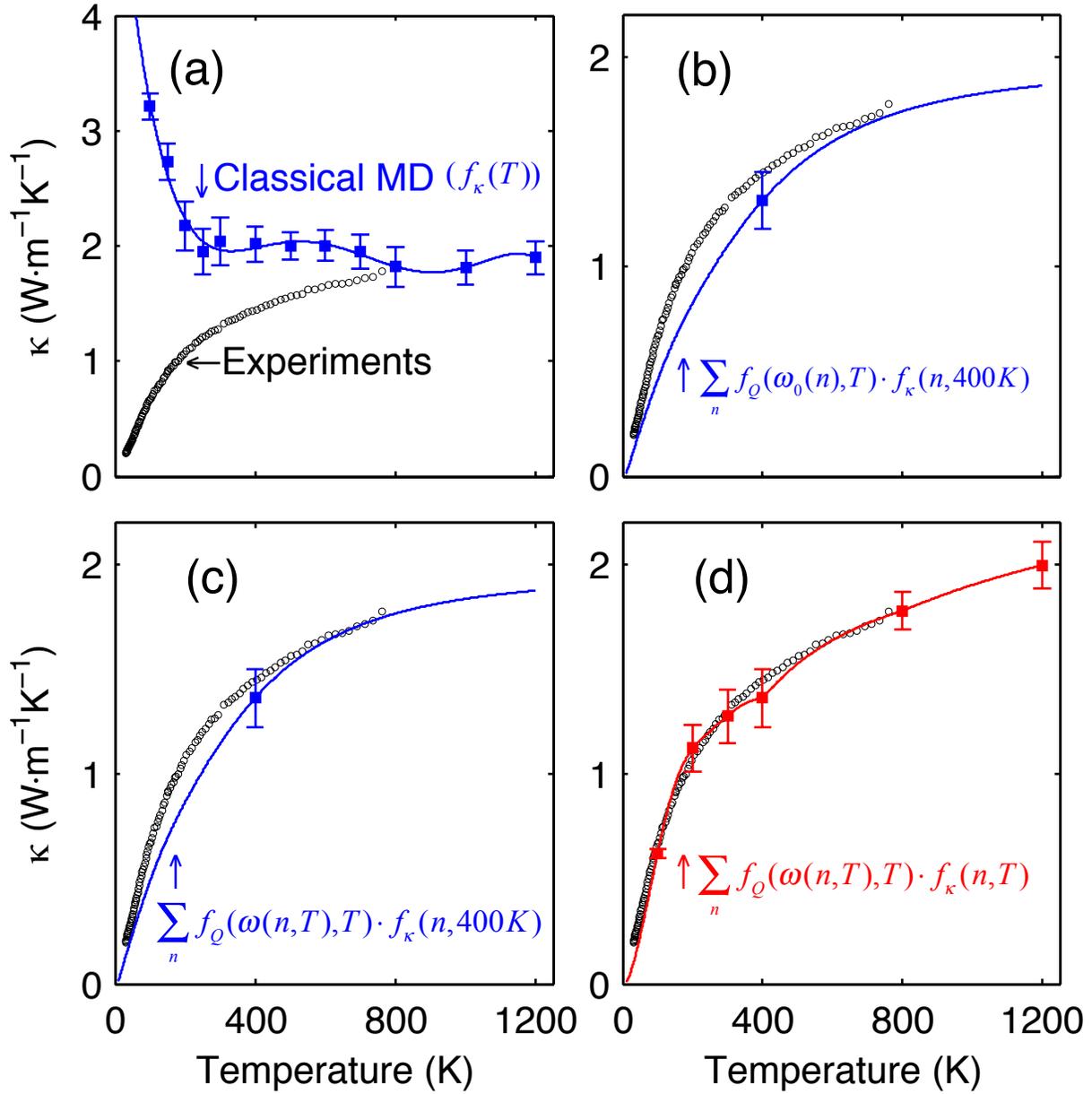

FIG. 2 (a) shows the GK results with error bars comparing with experiments (black circles); (b) is the result by using 400K GKMA data with quantum specific heat correction; (c) is based on previous result and including temperature dependent frequencies at 400K; (d) is the result using GKMA results at 100K, 200K, 300K, 400K, 800K and 1200K (interpolated in between) with the quantum specific heat correction and temperature dependent frequencies.

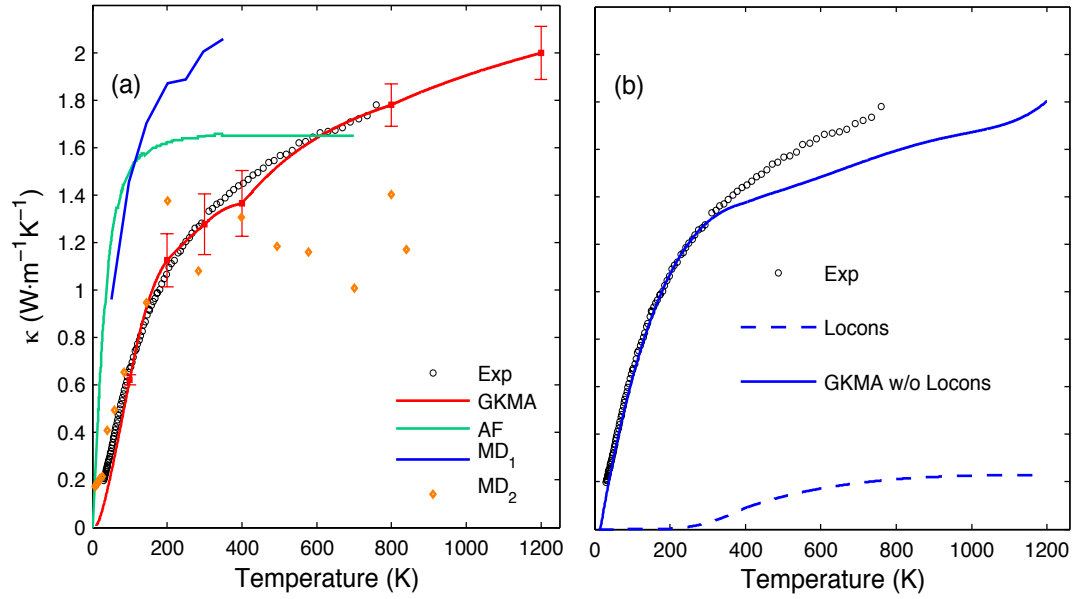

FIG. 3. (a) The red curve is temperature dependent TC of a-SiO$_2$ from GKMA calculations with uncertainty, experimental results (black circles) is from reference [27], AF is Allen-Feldman theory prediction from reference [26], MD$_1$ is non-equilibrium molecular dynamics simulation result from Shenogin et al. [26] and MD$_2$ is the MD result with quantum corrections from Jund and Jullien [30]. (b) The blue dashed curve represents the locons' contributions to the TC and solid blue curve is the sum of the delocalized modes contributions.

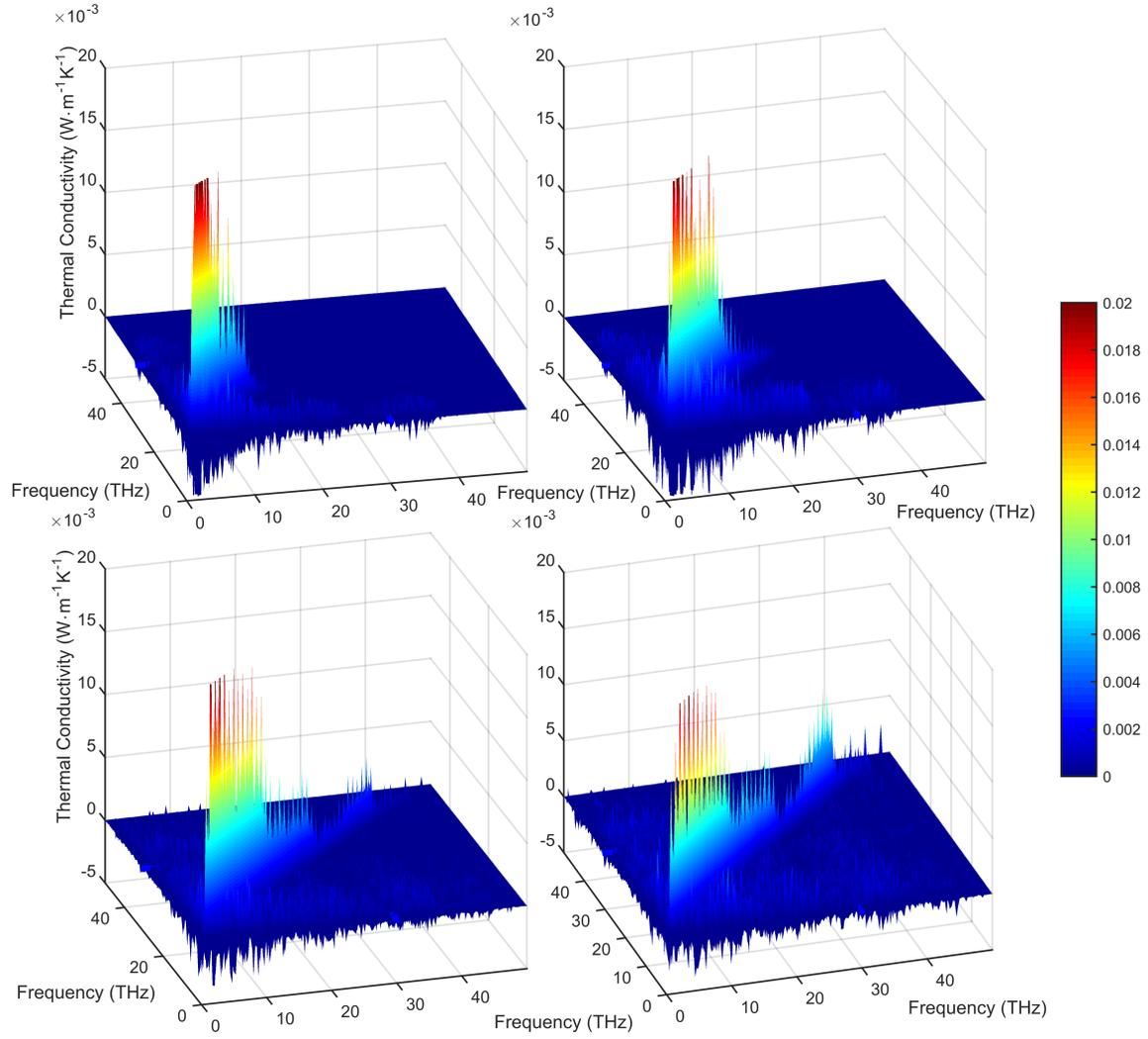

FIG. 4. (a) Cross-correlation map of thermal conductivity contributions including the quantum specific heat correction (detailed equations are shown in supplementary materials). The values are determined from the mode-mode cross-correlations for a-SiO$_2$ at 100K, (b) at 200K, (c) at 400K and (d) at 800K.